\documentclass{aa}  

\usepackage{graphicx}
\usepackage{txfonts}
\bibpunct[ ]{(}{)}{;}{a}{}{,}
\DeclareMathAlphabet{\mathsc}{OT1}{cmr}{m}{sc}
\def\testbx{bx}%
\DeclareRobustCommand{\ion}[2]{%
\relax\ifmmode
\ifx\testbx\f@series
{\mathbf{#1\,\mathsc{#2}}}\else
{\mathrm{#1\,\mathsc{#2}}}\fi
\else\textup{#1\,{\mdseries\textsc{#2}}}%
\fi}

\newcommand{\Ha}{H$\alpha$}

\newcommand{\kms}{km~s\ensuremath{^{-1}}}

\newcommand{\bv}{\ensuremath{B\!-\!V}}

\def\lya{\ensuremath{{\rm Ly}\alpha}}

\def\kms{km\,s$^{-1}$}

\newcommand{\lapp}{\mbox{\raisebox{-0.3em}{$\stackrel{\textstyle <}{\sim}$}}}
\newcommand{\gapp}{\mbox{\raisebox{-0.3em}{$\stackrel{\textstyle >}{\sim}$}}}
\newcommand{\be}{\begin{equation}}
\newcommand{\en}{\end{equation}}

\newcommand{\zabs}{\ensuremath{z_{\rm abs}}}
\newcommand{\zem}{\ensuremath{z_{\rm em}}}
\newcommand{\zq}{\ensuremath{z_{\rm {qso}}}}
\newcommand{\zg}{\ensuremath{z_{\rm {gal}}}}

\def\lya{Ly$\alpha$ }

\def\hi{H\,{\sc i}}
\def\mgii{Mg\,{\sc ii} }

\def\kms{km\,s$^{-1}$}
\def\Ha{\ensuremath{{\rm H}\,\alpha}}
\def\Hb{\ensuremath{{\rm H}\,\beta}}

\begin{document}

\title{21-cm absorption from galaxies at $z\sim0.3$}
\author{N. Gupta \inst{1, 2}
      \and R. Srianand \inst{2}
      \and P. Noterdaeme \inst{3} 
      \and P. Petitjean \inst{3} 
      \and S. Muzahid \inst{2} 
}

\institute{ASTRON, the Netherlands Institute for Radio Astronomy, Postbus 2, 7990 AA, Dwingeloo, The Netherlands
\\ \email{gupta@astron.nl} 
\and 
IUCAA, Post Bag 4, Ganeshkhind, Pune 411007, India
\and 
UPMC-CNRS, UMR7095, Institut d'Astrophysique de Paris, F-75014 Paris, France
}

\date{Received 30 Mar 2013; Accepted 16 Aug 2013}

\titlerunning{21-cm absorption from galaxies at $z\sim0.3$}

\keywords{
quasars: absorption lines -- 
galaxies: evolution --
galaxies: ISM --
galaxies: star formation.
}


\abstract{
We report the detection of 21-cm absorption from foreground galaxies 
towards quasars, specifically \zg=0.3120 towards SDSS J084957.97+510829.0 
(\zq=0.584; Pair-I) and \zg=0.3714 towards SDSS J144304.53+021419.3 
(\zq=1.82; Pair-II).
In both the cases, 
the integrated 21-cm optical depth is consistent with the absorbing
gas being a damped Lyman-$\alpha$ (DLA) system. 
In the case of Pair-I, strong Na~{\sc i} and Ca~{\sc ii} absorption lines are 
also detected at \zg\ in the QSO spectrum.  
We identify an early-type galaxy at an impact parameter of $b\sim$14\,kpc 
whose photometric redshift is consistent with that of the detected
metal and 21-cm absorption lines. This would be the first example 
of an early-type galaxy associated with an intervening 21-cm absorber.
The gas detected in 21-cm and metal absorption lines 
on the outskirts of this luminous red galaxy could be associated with 
the reservoir of cold \hi\ gas with a low level of star formation activity 
in the outer regions of the galaxy as reported in the literature for 
$z\sim$0.1 early-type galaxies. 
In the case of Pair-II, the absorption is associated with a low surface brightness galaxy  
that, unlike most other known quasar-galaxy pairs (QGPs), i.e. QSO sight lines passing 
through disks or halos of foreground galaxies, is identified only via 
narrow optical emission lines detected on top of the QSO spectra.  Using SDSS spectra we 
infer that the emission lines originate within $\sim$5\,kpc of the QSO sight
line, and the gas has metallicity [12+O/H$]\sim$8.4 and star formation rate$\sim$0.7-0.8\,M$_\odot$ yr$^{-1}$. 
The measured 21-cm optical depth can be reconciled with the $N$(H~{\sc i})
we derive from the measured extinction (A$_V$=0.6) if either the H~{\sc i} gas is warm or the extinction per 
hydrogen atom in this galaxy is much higher than the mean value of the Small Magellanic Cloud.
Finally, using a sample of 9 QGPs with 21-cm absorption detection from our observations and literature, 
we report a weak anti-correlation (Spearman rank, $r_s$=-0.3) between the 21-cm optical depth and galaxy impact 
parameter.  Milliarcsecond scale images and spectra are required to understand the implications  
of this.
}

\maketitle

\section{Introduction}

It is well known that physical conditions in the diffuse 
interstellar medium (ISM) of galaxies are influenced by 
various radiative and mechanical feedback processes associated with  
in-situ star formation \citep[e.g.,][]{Wolfire03}. 
Therefore, volume-filling factors of different phases of 
gas in a galaxy are expected to depend on its star formation history.
Of particular interest is the evolution of the volume-filling factor of 
cold neutral medium (CNM) phase that also serves as a gaseous reservoir 
for star formation in galaxies.  It is expected to contain an imprint of 
the collective outcome of all the processes that shape 
the star formation history of the Universe \citep[e.g.,][]{Hopkins06}.  
Systematic searches of high-$z$ {\it intervening} 21-cm absorbers in  
samples of Mg~{\sc ii} systems and damped \lya\ systems (DLAs) towards 
QSOs to measure CNM filling factor of galaxies have resulted
in detections of 21-cm absorption towards $\sim$10-20\% of Mg~{\sc ii} systems
\citep[e.g.,][]{Briggs83, Lane_phd, Gupta09, Kanekar09mg2, Gupta12} and DLAs 
\citep[e.g.,][]{Kanekar03, Curran10, Srianand12dla}.
However, establishing a connection between the redshift evolution
of 21-cm absorbers and global star formation rate density is not
straight forward due to (i) small number statistics of 21-cm
absorbers, (ii) ambiguities regarding the origin of absorbing gas
(i.e., gaseous disk, halo or outflowing/infalling gas etc.)
and (iii) issues related to the small scale structure in absorbing
gas and the extent of radio source. 

Blind searches of 21-cm absorption with Square Kilometre Array (SKA) 
pathfinders in the near future are expected to significantly increase the 
number of intervening 21-cm absorbers over a wide redshift range.
To understand the nature of absorbers detected from these, 
it is important to establish a link between the properties of 
galaxies and intervening 21-cm absorbers, and 
address issues related to points (ii) and (iii) mentioned above.   
At present, both these points are best addressed at $z\lapp0.2$ where 
it is possible to determine properties of absorbing gas on pc and kpc 
scales via arcsecond and milliarcsecond scale spectroscopy, and detect 
the galaxy responsible for absorption in 21-cm and various optical emission lines. 
However, almost all the intervening 21-cm absorption line searches until now 
have happened at $z\gapp$0.2.  
The notable exceptions in the recent past are \citet{Gupta10}, \citet{Borthakur11},  
and \citet{Darling11} \citep[see also][]{Carilli92}.  
The first two are based on samples of QSO sight lines passing through disks or halos of 
foreground galaxies, which we henceforth refer to as quasar-galaxy pairs (QGPs), whereas  
the last is a blind search for 21-cm absorption at $z\lapp0.06$ using the Arecibo telescope.         
All three have detected one 21-cm absorber each and showcase difficulties in 
detecting 21-cm absorbers at low-$z$ in the absence of large catalogs of Mg~{\sc ii} systems 
and DLAs.

To move forward, we are systematically searching the Sloan Digital Sky Survey (SDSS) 
for QGPs with angular separation $<$10$^{\prime\prime}$ and background quasar flux 
density $\gapp$100\,mJy in the Faint Images of the Radio Sky at Twenty-Centimeters (FIRST) 
catalog.  
The objective is to search for 21-cm absorption in QGPs with impact parameter $<$30\,kpc 
and systematically increase the number of low-$z$ intervening 21-cm absorbers/DLAs.  
This sample of low-$z$ absorbers can then be used as a comparison sample to understand 
the nature of high-$z$ 21-cm absorbers and DLAs \citep[see][and references therein for 
the current status of $z>2$ DLA host galaxies]{Krogager12}. 
In this paper, we report the detection of 21-cm absorption 
from two foreground galaxies at \zg$\sim$0.3 towards quasars 
SDSS\,J084957.97+510829.0 
(\zq=0.584, hereafter J0849+5108) and SDSS\,J144304.53+021419.3 (\zq=1.82, hereafter J1443+0214).   
These two pairs with foreground galaxies at $\zg\sim0.3$ were   
identified while scanning the SDSS images and spectra of bright radio sources 
and are interesting in their own right. 
In both these cases, the measured 21-cm optical depth is consistent with the 
absorbing gas being a DLA.    
The nature of background radio source in the case of QGP J0849+5108 
has been the subject of considerable debates and is suspected of being 
affected by foreground galaxy lensing. The quasar 
sight line is positioned close to a pair of low-$z$ ($\zg$$\sim$0.07) interacting spiral galaxies,   
and the field contains many bright galaxies within 60$^{\prime\prime}$ of the quasar 
\citep{Stickel89}.  The sight line has been previously searched for    
21-cm and molecular absorption lines \citep{Boisse88, Wiklind95, Gupta10}.  

The QGP J1443+0214 differs in the way it has been identified. 
Unlike most other QGPs studied to date, this QGP is identified by the 
novel technique of detecting emission lines from a foreground galaxy 
in the spectra of QSOs \citep[see e.g.,][]{Noterdaeme10o3, York12}.  
The foreground galaxy is otherwise invisible in this case. 
In Section~\ref{sec:obs}, we present details of our observations. 
We then detail the properties of individual QGPs in Section~\ref{sec:prop}. 
In Section~\ref{sec:disc}, we discuss implications of our results.
Throughout this paper we use the $\Lambda$CDM cosmology with 
$\Omega_m$=0.27, $\Omega_\Lambda$=0.73, and H$_{\rm o}$=71\,\kms\,Mpc$^{-1}$.

\section{Radio observations}
\label{sec:obs}

\begin{figure}
\centering
\hbox{
\includegraphics[width=85mm,height=80mm]{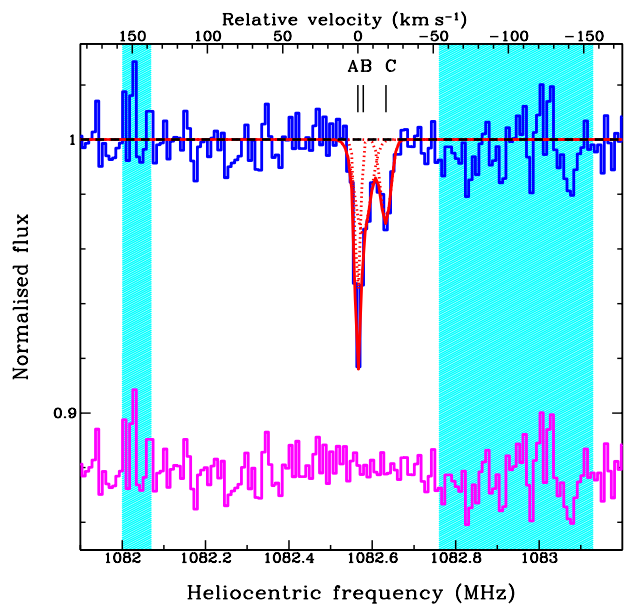}
}
\caption{
The 21-cm absorption detected towards the QGP J0849+5108. 
Origin of the velocity scale given at the top is defined with respect to the  maximum of optical depth.  
Individual Gaussian components labelled as A, B, and C (see Table~\ref{21cmfit}) and resulting fits to the absorption
profile are plotted as dotted and continuous lines, respectively. Residuals
are also shown with an arbitrary flux offset for clarity. Shaded regions 
mark the frequency range affected by RFI.
}
\label{j0849fit1}
\end{figure}

\begin{figure}
\centering
\hbox{
\hspace{0.0cm}
\includegraphics[width=85mm,height=80mm]{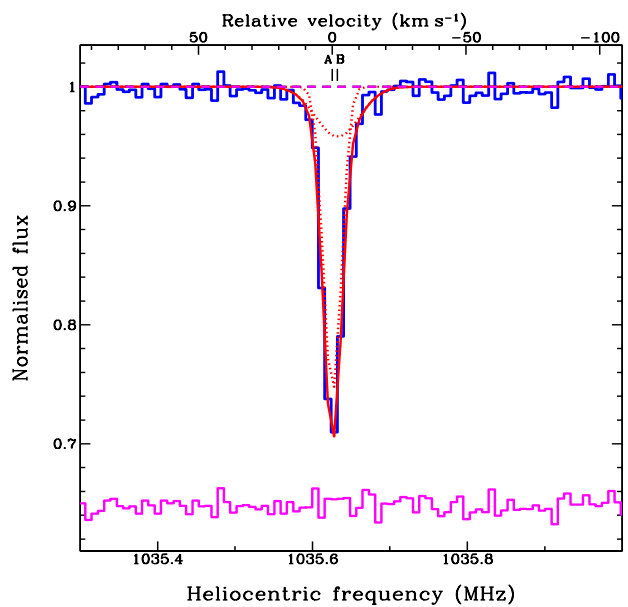}
}
\caption{The 21-cm absorption detected towards J1443+0214. 
The other details are same as in Fig.~\ref{j0849fit1}. 
}
\label{j1443fit}
\end{figure}

Both the pairs were observed with Giant Metrewave Radio Telescope (GMRT) using a baseband bandwidth 
of 4.17\,MHz split into 512 spectral channels (resolution$\sim$2.3\,\kms) 
centered on the redshifted 21-cm frequency of foreground galaxies.  
The QGP J0849+5108 was observed on 2011 July 18 (2.7\,hrs on source)   
and J1443+0214 on 2012 June 2 (5.7\,hrs on source).  
Standard calibrators were regularly observed during the observations for 
flux density, bandpass, and phase calibrations.  The data were reduced using 
the Astronomical Image Processing System (AIPS)  
following standard procedures as in \citet{Gupta10}. 
Both the quasars are represented well by a single Gaussian component,  
i.e. unresolved in our GMRT images that have spatial resolution 
of $\sim$4$^{\prime\prime}\times$3$^{\prime\prime}$.
The flux densities of J0849+5108 and J1443+0214 were measured to be 245\,mJy and 
163\,mJy, and spectral rms to be $\sim$2.0\,mJy\,beam$^{-1}$\,channel$^{-1}$ 
and $\sim$0.9\,mJy\,beam$^{-1}$\,channel$^{-1}$, respectively. 
In both cases, we detected 21-cm absorption consistent with the redshift 
of the foreground galaxy.   
In the case of J0849+5108, a few frequency channels on either 
side of the detected absorption feature were found to be affected 
by RFI (see shaded regions in Fig.~\ref{j0849fit1}).  
We therefore reobserved this QGP on 2012 July 3 (3.5\,hrs on source) to confirm the absorption. 
The flux density of the quasar was measured to be 221\,mJy, i.e. $\sim$10\% lower than the value  
measured in the first observing run.   
The velocity shift due to heliocentric motion of the Earth between the two 
observing runs on this QGP is 5.6\,\kms.  
The repeat observation reproduced the previously detected absorption feature 
at the expected frequency.  This confirms the reality of the absorption.
The spectrum presented in Fig.~\ref{j0849fit1} is obtained by the inverse-rms-square weighted 
average of the {\it normalized} spectra obtained from two observing runs.  
The 21-cm absorption spectrum towards J1443+0214 is presented in Fig.~\ref{j1443fit}.

In the case of absorber towards J0849+5108, the total integrated 21-cm optical depth, $\int\tau$$dv$= 
0.95$\pm$0.06\,\kms. Overall, 90\% of the total optical depth is contained within 22.5\,\kms. 
In the case of J1443+0214, the total integrated 21-cm optical depth is 3.4$\pm$0.1\,\kms, and 
90\% of the total optical depth is contained within 16.5\,\kms.  
We parametrize both the absorption profiles using multiple Gaussian components.  
The fits are shown in Figs.~\ref{j0849fit1} and \ref{j1443fit}, and details of  
Gaussian components are provided in Table~\ref{21cmfit}. 
\begin{table}[h]
\caption{Multiple Gaussian fits to the 21-cm absorption profiles. Listed are the component 
identification and corresponding redshift, full width at half maximum (FWHM), and peak optical depth. 
}
\begin{center}
\begin{tabular}{ccccc}
\hline
{\large \strut} Quasar        &  ID  &   \zabs           &            FWHM            &  $\tau_{peak}$ \\
                              &      &                   & (kms$^{-1}$)\\
\hline                                
{\large \strut} J0849+5108    &  A   &   0.312074        &            4$\pm$1         &     0.059$\pm$0.010   \\
                              &  B   &   0.312058        &           13$\pm$2         &     0.035$\pm$0.008   \\
                              &  C   &   0.311992        &            8$\pm$2         &     0.031$\pm$0.004   \\
{\large \strut} J1443+0214    &  A   &   0.371544        &            8$\pm$1         &     0.297$\pm$0.017   \\
                              &  B   &   0.371535        &           19$\pm$4         &     0.043$\pm$0.015   \\
\hline
\end{tabular}
\end{center}
\label{21cmfit}
\end{table}


\section{Physical properties in individual cases}
\label{sec:prop}

\subsection{\zg = 0.3120 galaxy towards J0849+5108}

\begin{figure}
\centering
\hbox{
\includegraphics[width=85mm,height=80mm]{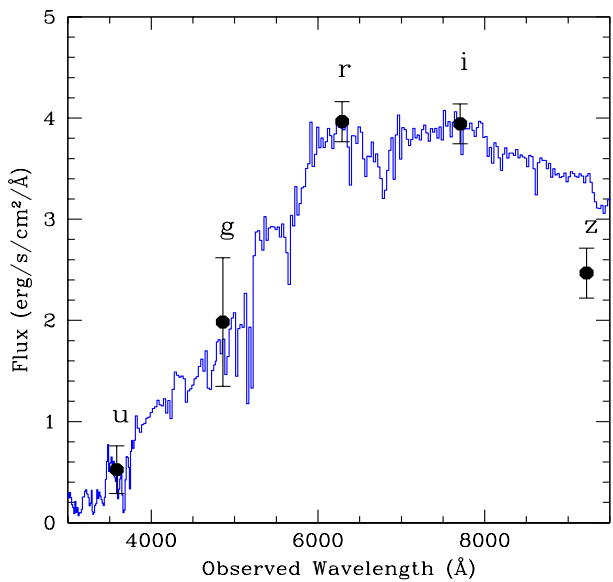}
}
\caption{
SDSS photometric points of the galaxy J084958.09+510826.7
overlaid on a typical luminous red galaxy spectrum at \zg=0.3120.
}
\label{j0849fit2}
\end{figure}

The background radio source in this case is a well-known object that has a BL Lacertae-like spectrum 
during outbursts \citep{Arp79}.  The nature of this object has been the subject of considerable debate. 
There are speculations about its optical spectrum being 
affected by foreground galaxy lensing and/or reddening \citep{Stickel89, Ostman06}.  
Interestingly, there are two foreground galaxies at angular separation less than 15$^{\prime\prime}$ 
to the radio source \citep[see Fig.~7 of][]{Gupta10}, and the sight line has been previously 
searched for 21-cm and molecular absorption \citep{Boisse88, Wiklind95}. 
The galaxy J084957.48+510842.3 (\zg=0.073) situated at $\sim$14$^{\prime\prime}$ ($b$$\sim$19\,kpc) 
north of the radio source is a member of an interacting pair of galaxies \citep{Stickel89}.  
No 21-cm absorption was detected from this galaxy by \citet{Boisse88}.  
The 3$\sigma$ 21-cm optical depth limit, $\int\tau_{3\sigma}dv$, was 0.32\,\kms. 
A deeper 21-cm absorption optical depth limit, $\int\tau_{3\sigma}dv$=0.08\,\kms, 
was obtained for this galaxy using GMRT by \citet{Gupta10}.
They also reported the detection of Na~{\sc i} and Ca~{\sc ii} 
absorption at \zg = 0.3120  in the spectrum of the QSO  
and associated it to a luminous red galaxy (LRG), identified as `G1' in their Fig.~7, 
at an angular separation of 2.3$^{\prime\prime}$ ($b$$\sim$14\,kpc) southeast of the QSO.  
While the spectrum of this galaxy is not available in the SDSS catalog,  
Fig.~\ref{j0849fit2} shows that the SDSS photometry is consistent with an LRG at the absorber's redshift. 
Here, we report the detection of 21-cm absorption at the redshift 
of Na~{\sc i} and Ca~{\sc ii} absorption lines (see Fig.~\ref{j0849fit1}).  
The 21-cm absorption profile is well fitted with three Gaussian components presented in 
Table~\ref{21cmfit}.  
In principle, the width of 21-cm absorption line can be used to constrain the kinetic temperature, 
$T_K\le$21.86$\times$FWHM$^2$, of absorbing gas \citep[e.g.,][]{Heiles03}.  
For J0849+5108, the FWHM of the strongest and narrowest Gaussian component `A' 
implies kinetic temperature, $T_K<$350\,K.   

The observed equivalent widths of Na~{\sc i} absorption lines correspond to 
$N$(Na~{\sc i})$\ge$ (6$\pm$2)$\times 10^{12}$ cm$^{-2}$, assuming the optically thin case. 
Consequently, we derive log\,$N$(\hi)$\ge21.2\pm0.5$ following the relation 
\begin{equation}
log\,\frac{N(\ion{Na}{i})}{N(\ion{H}{i})} = -(0.16\pm0.06)\,[log\,N(\ion{H}{i}) - 19.5] - (8.12\pm0.79)  
\end{equation}
from the known correlation between $N$(H~{\sc i}) and $N$(Na~{\sc i}) 
found in our Galaxy \citep{Ferlet85, Wakker00}.
By assuming an optically thin cloud with spin temperature $T_s$ and covering factor $f_c$, 
the total integrated 21-cm optical depth estimated from the GMRT spectrum 
corresponds to the \hi\ column density, 
$N$(H~{\sc i})=1.7$\times$10$^{20}$($T_s$/100)(1.0/$f_c$)\,cm$^{-2}$. 
Therefore, if the absorber was to follow the $N$(\hi)-$N$(Na~{\sc i}) relation seen in 
our galaxy, the harmonic mean spin temperature $T_s>$\,$890^{+1560}_{-580}$\,K.
Since such high $T_s$ values are typically seen in high-$z$
DLAs \citep{Kanekar03, Srianand12dla}, it will be important to have 
a direct Ly$\alpha$ measurement of $N$(H~{\sc i}) for this system .

The extent of gas detected in absorption is determined by the size of the background source.  
At milliarcsecond (mas) scales, J0849+5108 exhibits a core-jet morphology with an overall 
separation of $\sim$4\,mas, i.e. $\sim$20\,pc at the \zg\ \citep{D'ammando12}. 
The 5\,GHz Very Long Baseline Array (VLBA) observations simultaneous with the 
Very Large Array (VLA) observations recover 75\% to 85\% of the total 
arcsecond scale flux \citep[see Tables~6 and 7 of][]{D'ammando12}.  
The remaining flux originates in structures on scales $>$40\,mas (i.e., 200\,pc at \zg).    
This suggests that the region probed in the GMRT spectrum probably corresponds to   
scales $<$20\,pc.  If the extent of the absorbing gas is $>$20\,pc then this would 
imply $f_c>$0.7. 
Actually, the `core' component is unresolved with size$<$0.3\,mas (i.e., 1.4\,pc at \zg=0.3120) and 
contains about 95\% of the total mas-scale flux at 5-10\,GHz.  
Thus, the derived constraints on $f_c$ and extent of the region probed in 21-cm 
absorption are conservative.  

As previously mentioned, while J0849+5108 has a spectral energy distribution (SED) similar to that 
of a blazar, \citet{Yuan08} identified it as a radio-bright narrow-line 
Seyfert 1 AGN with an additional contribution to the optical flux coming from a 
mildly relativistic jet. The intrinsic spectrum is uncertain and most likely  
very different from the median QSO spectrum of \citet{VandenBerk01}.  
Therefore, unlike the case of QGP J1443+0214 discussed next,  
it is not possible to obtain an independent estimate of $N$(\hi) by SED fitting 
in this case.

\subsection{\zg=0.3714 galaxy towards J1443+0214}
\begin{figure}[!t]
\centering
\renewcommand{\tabcolsep}{1pt}
\begin{tabular}{cc}
\includegraphics[bb=175 341 394 628,clip=,angle=90, height=0.39\hsize]{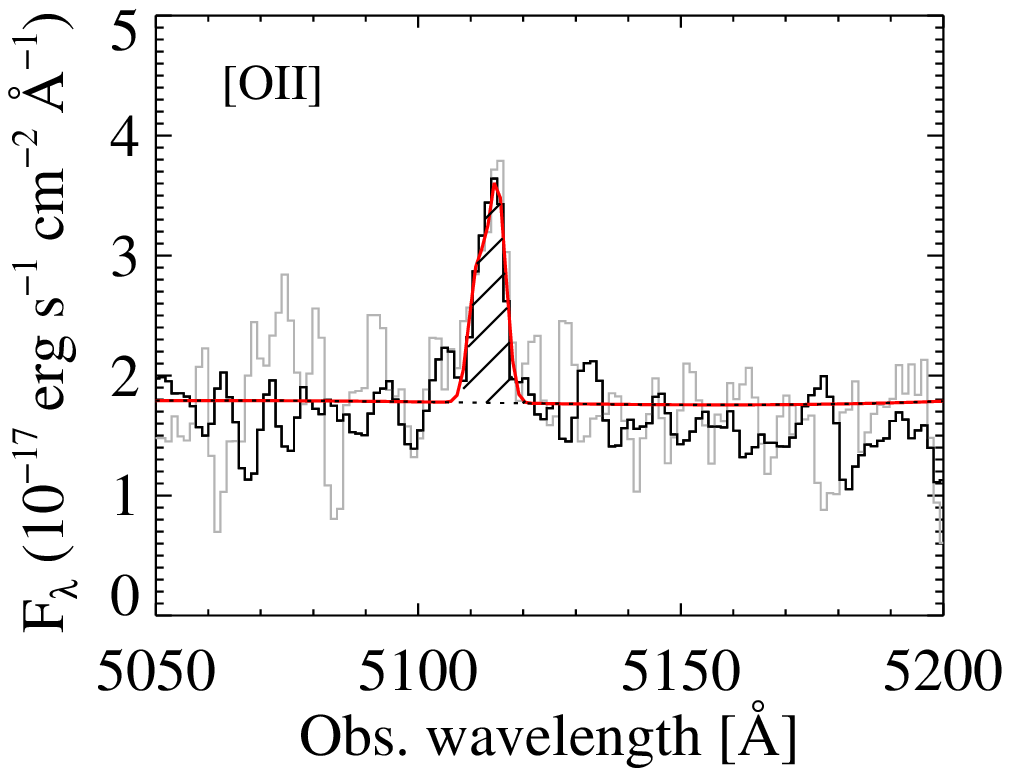} &
\includegraphics[bb=175 341 394 604,clip=,angle=90, height=0.39\hsize]{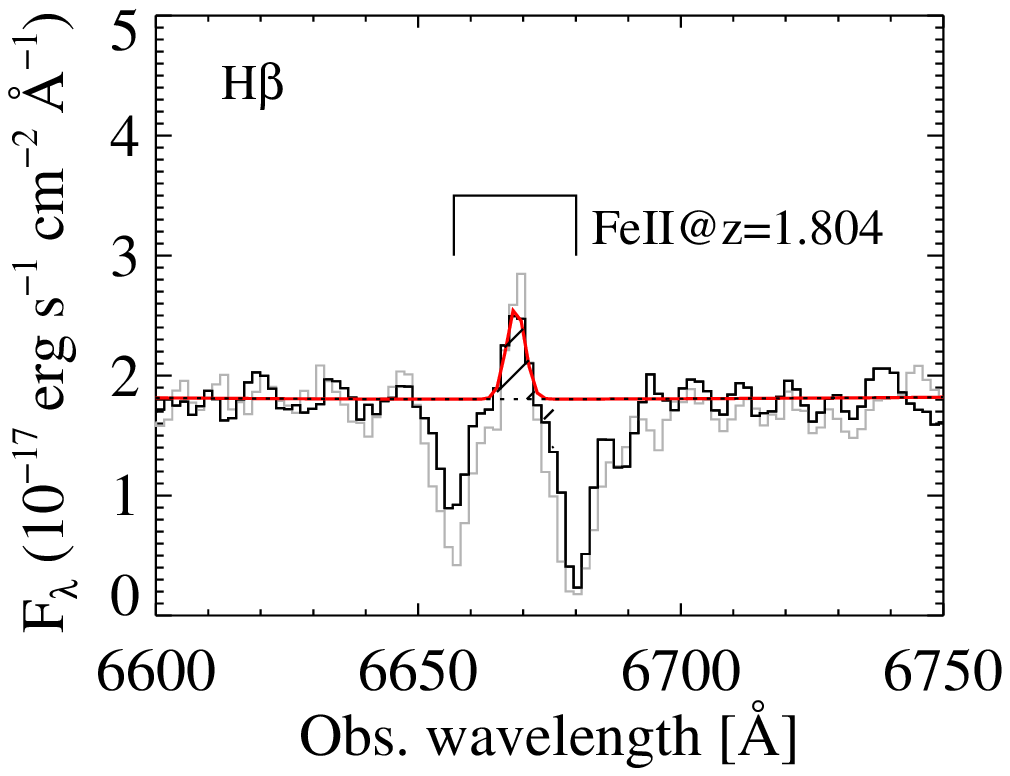} \\
\includegraphics[bb=175 341 394 628,clip=,angle=90, height=0.39\hsize]{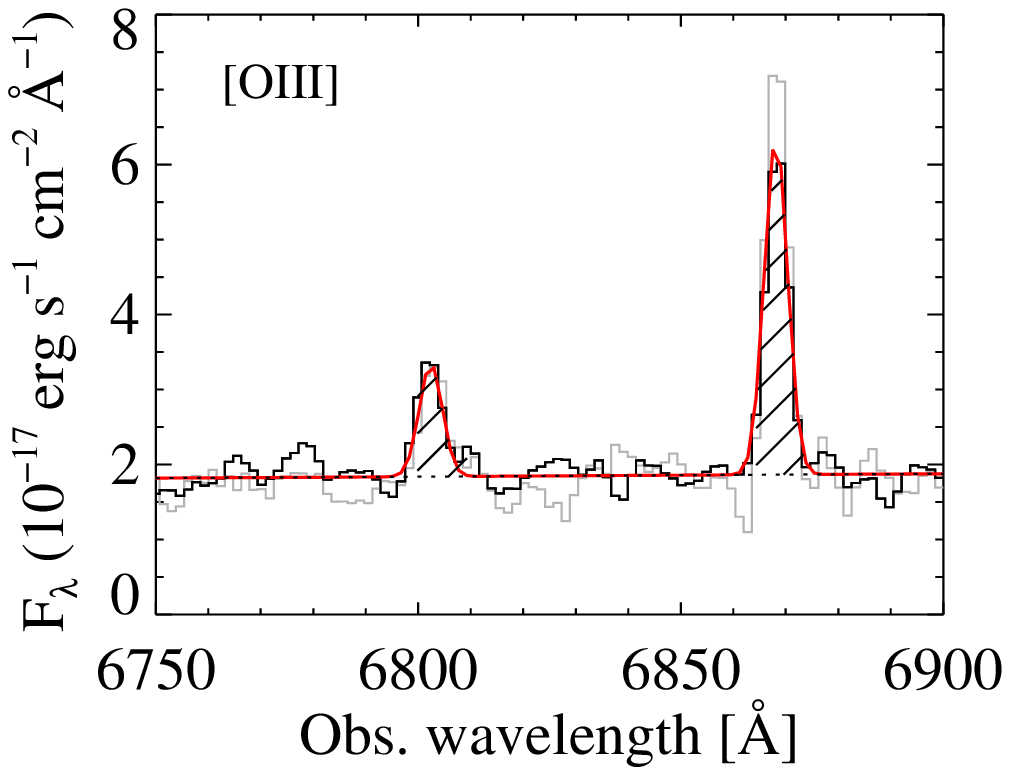} &
\includegraphics[bb=175 341 394 604,clip=,angle=90, height=0.39\hsize]{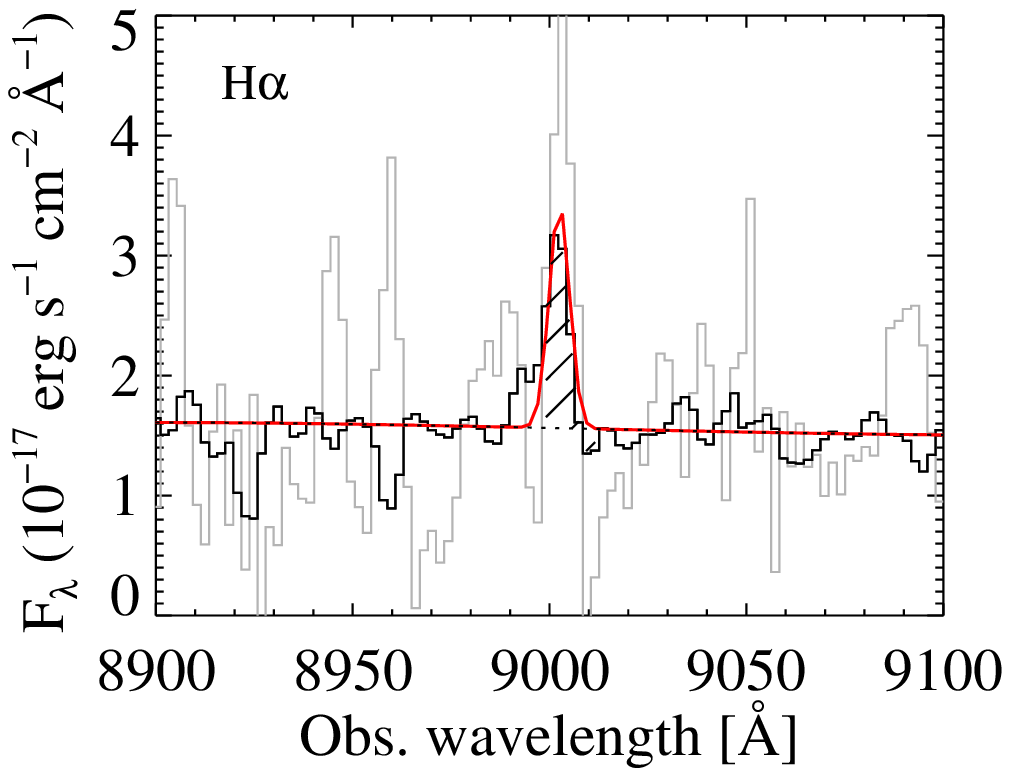} \\
\end{tabular}
\renewcommand{\tabcolsep}{6pt}
\caption{Fit to the galactic emission lines at $\zg=0.3714$ towards J1443+0214. The DR9 spectrum 
is shown in black, and the 3-pixel boxcar-smoothed DR7 spectrum is shown in gray.
The absorption lines seen on each side of H\,$\beta$ are due to 
Fe\,{\sc ii}\,$\lambda\lambda$2344,2374 at $\zabs=1.804$. 
\label{j1443_em}}
\end{figure}

Unlike most QGPs studied until now, where one clearly identifies a  
foreground galaxy in optical images, the galaxy in the case of QGP J1443+0214 is 
identified by the detection of emission lines in the SDSS spectrum of QSO.  
The foreground galaxy is invisible otherwise.
The SDSS spectrum [both in SDSS-II (DR7) and SDSS-III (DR9)] of QSO J1443+0214 
shows broad emission lines at \zem = 1.82 and a set of narrow emission 
lines (i.e., [O\,{\sc ii}], [O\,{\sc iii}], H\,$\beta$, and H\,$\alpha$) 
at \zg=0.3714 (see Fig.~\ref{j1443_em}) like the
foreground galaxies detected by \citet{Noterdaeme10o3} in the SDSS spectra of high-$z$ QSOs.
We do not detect any significant difference in the line strengths between the SDSS-II (DR7) and 
SDSS-III (DR9) spectra, while the fiber diameters are of 3\arcsec (15\,kpc at \zg) 
and 2\arcsec (10\,kpc at \zg), respectively. Therefore the line emission probably
originates in a low surface brightness region within 5\,kpc to the QSO sight line. 
This is also confirmed by the  non-detection of the galaxy in SDSS images after subtracting 
the QSO contribution and by the fact that the SED fitting as in Fig.~\ref{j1443red} (see below) did not 
require any additional contribution from the galaxy continuum.

We estimated the color excess E(B-V) by fitting the SDSS-II SED  
of QSO J1443+0214 by the SDSS-II QSO composite from \citet[][]{VandenBerk01}. 
The $<$2000\,\AA\ rest-wavelength coverage for the composite spectrum was taken 
from \citet[][]{Telfer02}.  We reddened the composite spectrum using the Galaxy, 
the LMC, and the SMC extinction curves \citep[see][for the detailed 
procedure]{Srianand08bump,Noterdaeme09co,Noterdaeme10co}. 
These extinction curves differ significantly only at $\lambda$$<$2500\,\AA. For J1443+0214, 
the SMC extinction curve is favored by the ultraviolet photometric points from the GALEX images.  
The J1443+0214 SED, along with the composite QSO SED reddened using the SMC extinction 
curve at $\zg = 0.3714$ with the best fit A$_V=0.61$, is shown in Fig.~\ref{j1443red}.
We went on to apply the same procedure to a control sample of 223 SDSS-II QSOs from 
\citet{Schneider10} with emission redshifts within $\Delta z = \pm 0.005$ from that of J1443+0214,  
where we removed QSOs with strong BAL activity using \citet{Allen11}. 
We estimated the uncertainty due to intrinsic shape variations from the width of A$_V$ distribution 
and corrected for the ``zero-point'' (the median A$_V$ value).
The extinction we thus derive for J1443 +0214 is A$_V=0.6\pm0.2$, i.e., $3\sigma$  
evidence of reddening along the sight line when compared to the control sample.  
Since we did not attempt to remove intervening absorbers from the control 
sample, the estimated A$_V$ and its statistical significance are conservative. 
%
\begin{figure*}[!ht]
\centering
\def\svgwidth{200pt}
\includegraphics[bb=110 22 500 760,clip=, angle=90, width=15cm]{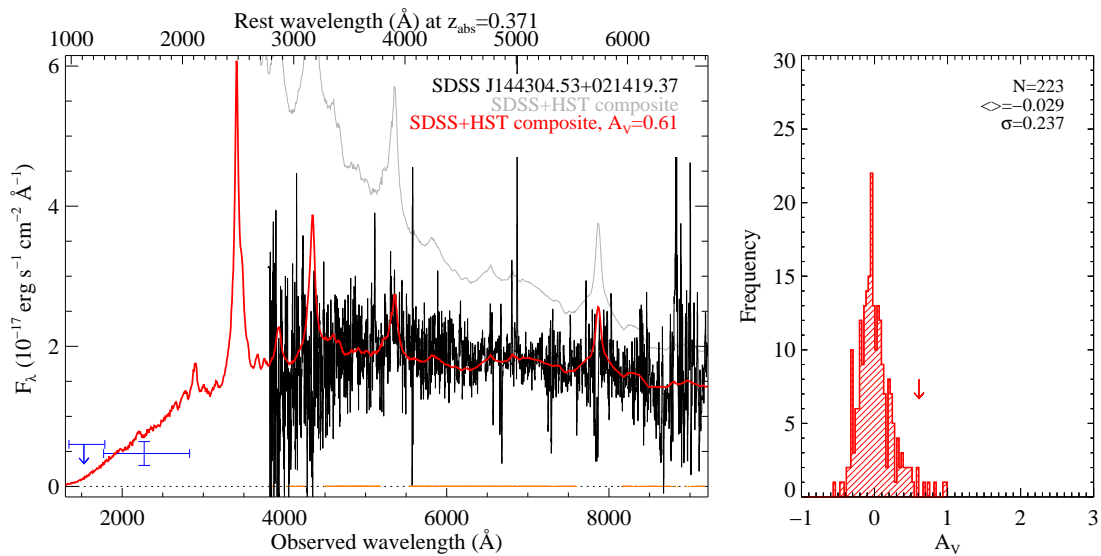}
\caption{{\it Left:} The SDSS spectrum of J1443+0214 (black) compared with the 
unreddened QSO composite (grey) and matched with the same QSO composite 
with  SMC-extinction law at $\zg = 0.3714$ and A$_V=0.61$ (red). 
In the ultraviolet range (points with error bars), we use the photometric data from GALEX to estimate
the flux.
The orange segments at y=0 indicate the spectral range considered in the fitting process. 
{\it Right:} The distribution of A$_V$ using SMC extinction law for a control sample of QSOs.  
The number of QSOs (N), the median ($<>$), and rms ($\sigma$) A$_V$ are also provided 
(see text for details).  The arrow indicates the position of QSO J1443+0214.
} 
\label{j1443red}
\end{figure*}

Next we use the Balmer decrement to obtain an independent estimate of the 
extinction.  Various emission line fluxes estimated through the Gaussian profile 
fitting using a higher signal-to-noise ratio DR9 spectrum \citep{Paris12} are given in 
Table~\ref{j1443_emt}.
As noted above, different extinction curves are identical over the wavelength 
range of emission lines considered here and therefore yield similar values.
From the observed Balmer decrement, we estimate the color excess 
$E(\bv) = E(\Hb\!-\!\Ha)/(k(\Hb)-k(\Ha)) \approx $ 0.19 and 0.20  
assuming intrinsic $\Ha/\Hb=2.86$ and $k(\lambda)$ for the 
SMC- and the Galaxy-type dust respectively \citep{Gordon03}.  
This corresponds to $A_V = 0.6$ for $R_V$ = 2.74 seen in the SMC. 
The extinction detected towards the QSO is significantly higher than the value 
(E(B-V)$<$0.03) generally seen towards intervening DLAs and \mgii\ systems 
\citep[e.g.,][]{York06, Khare12}. 
It is also interesting to note that the extinction derived in the star-forming region using
emission lines is remarkably consistent with the value derived along the QSO line of sight 
using SED fitting.  This suggests that the abundance of dust is homogeneous over the 
length scale corresponding to the impact parameter ($\sim$5\,kpc), over 
which the line fluxes are measured. Using the mean 
$A_V$ vs $N$(H~{\sc i}) relationship found for the SMC \citep{Gordon03}, we infer 
$N$(H~{\sc i}) = $(8\pm3)\times10^{21}$\,cm$^{-2}$.

\begin{table}[!ht]
\centering
\renewcommand{\tabcolsep}{2pt}
\caption{Emission line properties from the $\zg=0.3714$ galaxy towards J1443+0214 \label{j1443_emt}}
\begin{tabular}{c c c c}
\hline
\hline
{\large\strut} Line                          & $F_{\rm obs}$  & $L$\tablefootmark{(a)} & Derived \\
                                             & ($10^{-17}$~erg\,s$^{-1}$\,cm$^{-2}$) & ($10^{40}$~erg\,s$^{-1}$) & quantities \\
\hline
{\large \strut}$[$O\,{\sc ii}$]$             & 11.7 & 12.5 & SFR = 0.8~M$_{\odot}$\,yr$^{-1}$\tablefootmark{(b)}  \\
$[$O\,{\sc iii}$]\lambda$5007 & 24.8 & 21.4 & R$_{23}$ = 13.8                                        \\
H\,$\beta$                    &  3.4 & 3.0  & $E(\bv)$ = 0.20                                      \\
H\,$\alpha$                   & 12.1 & 8.5  & SFR = 0.7~M$_{\odot}$\,yr$^{-1}$\tablefootmark{(c)}   \\ 
\hline
\end{tabular}
\tablefoot{
\tablefoottext{a}{Luminosity corrected for dust extinction}
\tablefoottext{b}{$L_{\rm [OII]}$~-~SFR calibration from \citet{Kewley04}.}
\tablefoottext{c}{$L_{\rm H\,\alpha}$~-~SFR calibration from \citet{Kennicutt98}.}
}
\renewcommand{\tabcolsep}{6pt} 
\end{table}
%

%
The profile of the 21-cm absorption line detected towards J1443+0214 is relatively simple 
(Fig.~\ref{j1443fit}). In addition to a narrow component that can be modeled 
with a Gaussian of FWHM$\sim$8\,\kms\ ($T_K<$1400\,K), a broad component is also 
required to reasonably fit absorption in wings (Table~\ref{21cmfit}).
The total integrated 21-cm optical depth corresponds to  
$N$(H~{\sc i})=6.2$\times$10$^{20}$($T_s$/100)(1.0/$f_c$)\,cm$^{-2}$, 
which is less than the $N$(H~{\sc i}) we inferred from the SED fitting. 
This could either be due to the covering factor of the gas being much less 
than 1 or the harmonic mean $T_s$ being as high as $\sim$1400\,K.  
No mas-scale images or $N$(\hi) measurements using \lya are available 
to distinguish between these scenarios. 
Another possibility is that the extinction per hydrogen atom in this absorber 
is much higher than what is measured in the SMC.   
It has been suggested that this may be the case in some high-$z$ dusty 
Mg~{\sc ii} absorbers with 21-cm absorption \citep{Gupta12} and DLAs with CO detections \citep{Noterdaeme10co}.
 
\begin{figure}
\centering
\hbox{
\hspace{-0.5cm}
\includegraphics[width=100mm,height=85mm]{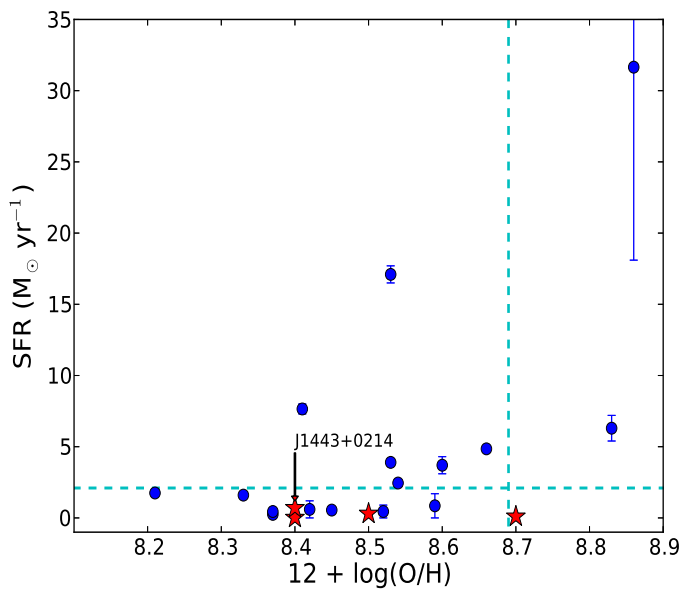}
}
\caption{Star formation rate versus metallicity for galaxies with 21-cm absorption 
(\zg$<$0.4; stars) and  [O~{\sc iii}]-emitting galaxies (0.4$<$\zg$<$0.8; circles) 
from \citet{Noterdaeme10o3}.  
The horizontal dashed line marks the median SFR for [O~{\sc iii}]-emitting galaxies. 
The vertical dashed line corresponds to solar abundance, 12 + log\,(O/H)$_\odot$ = 8.69 
\citep{Asplund09}.   
}
\label{sfr}
\end{figure}

Now we use nebular emission line properties to 
obtain insight into the nature of the absorbing galaxy.  
We correct emission line fluxes for extinction and use these to estimate the 
global metallicity and star formation rate (SFR) of the galaxy (Table~\ref{j1443_emt}).  
For J1443+0214 we estimate the R$_{23}$ ratio,  defined as 
([O\,{\sc ii}] + [O\,{\sc iii}])/H$\beta$, to be 13.8.  
In general, the R$_{23}$ calibration is double-valued with respect to metallicity, and 
additional line ratios are required to break this degeneracy. 
For J1443+0214, the measured ratio is a high value at the turnover of the high- and 
low-metallicity branches of R$_{23}$-metallicity calibration 
\citep[see e.g.,][]{Kobulnicky99,Pilyugin00}, 
indicating 12 + log\,(O/H)$\sim 8.4$ i.e., 0.5Z$_\odot$ using  
12 + log\,(O/H)$_\odot$ = 8.69 \citep{Asplund09}.  
From extinction-corrected \Ha\ luminosity, $L$(H$\alpha$), and using the 
following relation from \citet{Kennicutt98}, 
\begin{equation}
{\rm SFR(M_\odot\,yr^{-1})} = 7.9\times10^{-42} L({\rm H}\alpha)\,({\rm ergs\,s^{-1}}), 
\end{equation}
we derive SFR$\sim$0.7\,M$_{\odot}$\,yr$^{-1}$.  
Using the following relation from \citet{Kewley04} 
\begin{equation}
{\rm SFR(M_\odot\,yr^{-1})} = (6.6\pm1.7)\times10^{-42} L([{\rm O~{II}}])\,({\rm ergs\,s^{-1}}) 
\end{equation}
a similar value of 0.8$\pm$0.2~M$_{\odot}$\,yr$^{-1}$ is derived using L(O~{\sc ii}). 
Thus, the SFR is slightly below the median value, and the metallicity is well within 
the range of values found for such [O~{\sc iii}]-emitting galaxies by 
\citet{Noterdaeme10o3} (see Fig.~\ref{sfr}). 
In Fig.~\ref{sfr}, the metallicity measurements for galaxies detected by 
\citet{Noterdaeme10o3} correspond to the upper branch of R$_{23}$-metallicity 
calibration.  The corresponding metallicities on the lower R$_{23}$ branch 
would be typically 0.4\,dex lower.  The SFRs for these have been estimated using 
calibrations (Eqs. 23 and 24) from \citet{Argence09} that use uncorrected [O~{\sc ii}] and 
H$\beta$ luminosities. The error bars in Fig.~\ref{sfr} indicate the range of SFRs allowed by 
variations in dust attenuation and metallicity but do not account for fiber losses.

\begin{table*}[!ht]
\centering
\renewcommand{\tabcolsep}{2pt}
\caption{Emission line properties of 21-cm absorbers/DLAs  \label{21cmlit}}
\begin{tabular}{c c c c c c c c c c c}
\hline
\hline
{\large\strut}  Quasar     &   \zq   &    Galaxy      &   \zg       & $b$   & $\int\tau$dv &  12+log(O/H) &~~[O/H]\tablefootmark{(a)}  &SFR  & $\Sigma_{\rm SFR}$\tablefootmark{(b)}  & Reference \\
                           &         &                &             & (kpc) &   (\kms)     &   &       & (M$_{\odot}$\,yr$^{-1}$) & \\ 
\hline
{\large \strut} J104257+074850 & 2.7   & J104257+074751 & 0.03  & 1.7  & 0.2  &8.4 &$-$0.29 &0.01 &  ~~4.1    & \citet{Borthakur10} \\
                J124157+633241 & 2.6   & J124157+633237 & 0.14  & 11   & 2.9  &8.7 & ~~0.01 &0.1  &  ~~2.9    & \citet{Gupta10} \\
                J144304+021419 & 1.8   &     $-$        & 0.37  &$<$5  & 3.4  &8.4 & ~~0.29 &0.7  & $>$3.8    &  This work      \\
                J163956+112758 & 1.0   & J163956+112802 & 0.08  & 4.0  & 15.7 &8.5 &$-$0.19 &0.3  &  ~~18     & \citet{Srianand13dib} \\
\hline
\end{tabular}
\tablefoot{
\tablefoottext{a}{[O/H] = log(O/H) - log(O/H)$_\odot$}
 \tablefoottext{b}{in units of 10$^{-3}$M$_{\odot}$\,yr$^{-1}$\,kpc$^{-2}$}
}
\renewcommand{\tabcolsep}{6pt} 
\end{table*}

In addition to QGP J1443+0214, 
galactic emission lines superimposed on top of QSO continuum in SDSS fiber 
(diameter 3$^{\prime\prime}$) spectra have been detected for three other 21-cm absorbers. 
For these four absorbers, we are able to obtain SFR, SFR per unit area ($\Sigma_{\rm SFR}$) 
and metallicity (see Table~\ref{21cmlit} and references given in the last column for details).  
The SFR and metallicity for these 21-cm absorbers are also plotted in Fig.~\ref{sfr} for   
comparison with the sample of [O~{\sc iii}]-selected normal galaxies of \citet{Noterdaeme10o3}.     
Figure~\ref{sfr} suggests that the SFR of 21-cm absorbers is less than the median SFR 
of [O~{\sc iii}]-selected Mg~{\sc ii} absorbers.  While in the case of J1443+0214 we are 
sure that all the galactic emission is included in the SDSS fiber, in other cases a 
large fraction of foreground galaxy emission falls outside the SDSS fiber.  Therefore, 
SFRs for these galaxies are at best lower limits, and are in general underestimated with 
respect to the higher redshift (0.4$<\zg<$0.7) sample of \citet[][]{Noterdaeme10o3}.  
Therefore, within uncertainties SFRs for these QGPs with 21-cm absorption seem to be 
consistent with [O~{\sc iii}]-selected Mg~{\sc ii} absorbers.

\section[]{Discussion}
\label{sec:disc}

We have detected 21-cm absorption from two foreground galaxies at $z$$\sim$0.3 towards 
quasars J0849+5108 (b$\sim$14\,kpc) and J1443+0214 (b$<$5\,kpc).  
In both cases, we infer $N$(\hi)$\gapp$\,2$\times$10$^{20}$\,cm$^{-2}$, which qualifies 
these systems as DLAs. 
The integrated 21-cm optical depths are 0.95\,\kms\ for J0849+5108 and 3.4\,\kms\ for 
J1443+0214.  As can be noted from Fig.~19 of \citet{Gupta10}, such high values are observed 
among $z<1$ DLAs with host galaxy located at $b<$15\,kpc.   
In the case of J0849+5108, the SDSS photometry suggests that the Na~{\sc i}, Ca~{\sc ii}, and 21-cm 
absorption lines detected by us are associated with an LRG having a smooth optical morphology 
that is typical of a passively evolving early-type galaxy.  
In the case of J1443+0214, the absorption is probably associated with a low surface brightness (LSB) galaxy 
identified via optical emission lines within $\sim$5\,kpc of the QSO sight line.  
Late-type or LSB galaxies have been identified with a few low-$z$ DLAs and 21-cm absorbers 
\citep[see Table~4 of][]{Rao03, Lebrun97}.  However, the case of LRG associated with the absorber towards J0849+5108  
is the first example of an early-type galaxy associated with an intervening 21-cm absorber.  The gas 
detected in 21-cm and metal absorption lines in the outskirts (b$\sim$14\,kpc) of this LRG could 
be associated with the extended \hi\ disks ($N$(\hi)$\sim$10$^{20}$\,cm$^{-2}$) detected in deep 
\hi\ 21-cm emission observations of local early-type galaxies \citep[e.g.,][]{Serra12}. 
In particular, the cold gas detected in 21-cm absorption could be serving as fuel to the renewed 
star formation activity in the outer regions of the galaxy.   
This idea is supported by detections of ultraviolet rings that correspond to extended low-level 
recent or ongoing star formation in outskirts of $z\sim0.1$ early-type galaxies \citep{Salim12, Fang12}.

    
DLAs, as well as 21-cm absorption line surveys at $z<2$, are generally based on QSO 
sight lines selected via \mgii\ absorption \citep{Rao06}, and it is well known that  
LRGs offer very little cross-section to \mgii\ absorption \citep{Bowen11}.  
Indeed, in an Mg~{\sc ii} selected sample of DLAs, sub-DLAs and Lyman limit 
systems at 0.1$<z<$1.0 \citet{Rao11} find that absorber galaxies comprise a mix of spectral types 
that are predominantly late type.  Of eight DLA galaxies in the part of their sample with 
sufficient photometry to carry out galaxy template fits, only one is found to be early type.
Clearly, much larger samples are required to quantify the prevalence of early-type 
galaxies in samples of DLA and 21-cm absorber hosts.  
The techniques that rely solely on the detection of nebular emission lines to 
identify absorber hosts may miss such cases. 

Now we turn to low-$z$ 21-cm absorbers with detections of nebular emission lines 
in QSO spectra.  The SFR of absorbing galaxy in the case of J1443+0214 presented 
here is well below the median SFR found for [O~{\sc iii}]-selected Mg~{\sc ii} absorbers 
at 0.4$<z<$0.7 \citep{Noterdaeme10o3}. 
A large fraction (87\%) of these [O~{\sc iii}]-selected absorbers exhibits 
strong Mg~{\sc ii} absorption ($W_{\rm r}>$1\AA).  This, along with the detection of strong Fe~{\sc ii} 
absorption in several cases, suggests that most of these systems could be DLAs 
\citep[see][]{Rao06}.  
\citet{Wolfe03b} have used C~{\sc ii}$^*$ to infer $\Sigma_{\rm SFR}$ in  
$z>2$ DLAs.  For the CNM-model i.e. where QSO sight lines pass through  
CNM, they measure $\Sigma_{\rm SFR}$=10$^{-2.2}$\,M$_\odot$\,yr$^{-1}$\,kpc$^{-2}$ 
and measure $\Sigma_{\rm SFR}$=10$^{-1.3}$\,M$_\odot$\,yr$^{-1}$\,kpc$^{-2}$ for the 
warm neutral medium (WNM)-model.  The $\Sigma_{\rm SFR}$ that we measure for  
low-$z$ QGPs in Table~\ref{21cmlit} are more in line with the CNM-model.  This is 
not surprising since the presence of CNM in these cases is confirmed by the detection 
of 21-cm absorption.  
The abundances we infer for these QGPs using emission lines are in the range, 
[O/H]$\sim$ -0.3 to 0.3 (Table~\ref{21cmlit}), and are significantly higher than 
the values generally inferred for $z>2$ DLAs using absorption lines. 
It will be interesting to obtain ultraviolet spectra of 
these QGPs to measure $N$(\hi) and metallicity.  This will  
also provide a direct measurement of spin temperature and CNM fraction along the 
sight lines, and will be crucial for understanding low CNM fractions in $z>2$ DLAs 
inferred from the observations of 21-cm and H$_2$ absorption 
\citep[e.g.,][]{Kanekar03, Srianand12dla}.

\begin{figure}
\centering
\hbox{
\hspace{-0.5cm}
\includegraphics[width=100mm,height=85mm]{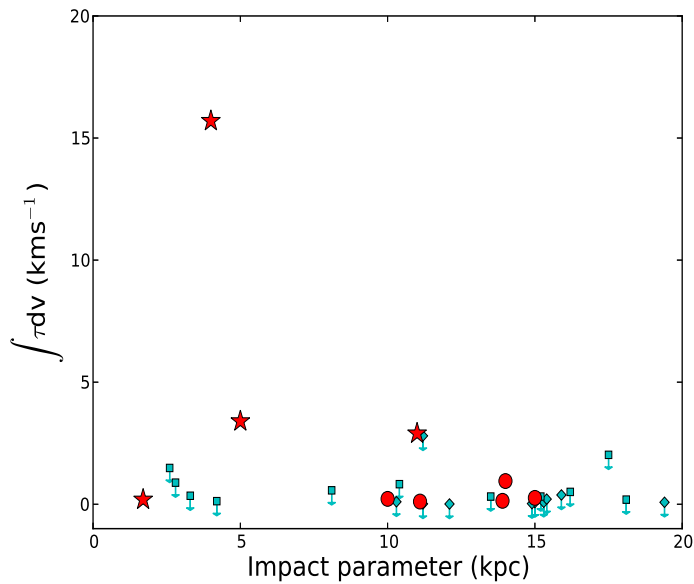}
}
\caption{Integrated 21-cm optical depth vs impact parameter for QGPs with 
21-cm absorption measurements.  The QGPs from Table~\ref{21cmlit} are plotted as stars.
Also plotted as diamonds and squares  are 3$\sigma$ upper limits 
corresponding to a velocity resolution of 10\,\kms\ 
from Tables~2 and 5 of \citet[][]{Gupta10} and Table~3 of \citet[][]{Borthakur11}, 
respectively. 
}
\label{imp}
\end{figure}
  
At $z>2$, where it is possible to build large samples of DLAs via ground-based 
optical surveys \citep{Noterdaeme12dla}, searches to detect DLA host galaxies have 
mostly resulted in non-detections \citep[e.g.,][]{Moller93, Rahmani10}.
Recently, thanks to powerful instrumentation on the VLT, detections of a few 
DLA host galaxies have been possible by specifically targeting high-metallicity/$N$(\hi) 
DLAs \citep[see e.g.,][]{Fynbo11, Noterdaeme12dlagal}. 
Using a sample of ten DLA hosts at $z>2$, \citet[][]{Krogager12} report anti-correlation 
(Spearman rank, $r_s$=$-$0.6) between log\,$N$(\hi) and the galaxy impact parameter \citep[see also][]{Peroux11}.  
A similar trend is expected between 21-cm optical depth and impact parameter albeit with a large 
scatter probably due to further dependence on $T_s$ and $f_c$.  The strength of this correlation can 
potentially shed light on processes driving the CNM filling factor of  
DLAs.  In Fig.~\ref{imp}, we plot 21-cm optical depth versus impact parameter 
for all the $z\lapp$0.4 QGPs with 21-cm absorption measurements \citep[][]{Gupta10, Borthakur11}. 
The Spearman rank correlation coefficient test suggests weak ($r_s$=-0.3) anti-correlation 
between $\int\tau$dv and $b$ for QGPs with 21-cm detection.  Milliarcsecond scale images and 21-cm absorption spectra 
along with a much larger sample, especially with $b<$10\,kpc, are required to confirm and 
understand the implication of this trend.     
Because the number of DLAs and consequently DLA hosts at $z>2$ is expected to steadily increase 
with the availability of large samples of QSO spectra from the SDSS Baryon Oscillation Spectroscopic 
Survey \citep[][]{Noterdaeme12dla}, all this clearly motivates the need for large surveys of 
21-cm absorption line especially at low-$z$ where it is relatively easy to identify and
determine the properties of absorbing galaxy.  
Such surveys will become possible in the near future with SKA pathfinders and will uncover 
a new population of high-$N$(\hi) absorbers i.e., DLA/sub-DLAs that have escaped optical/ultraviolet  
absorption line surveys due to biases caused by either dust extinction or pre-selection 
of QSO sight lines based on various metal absorption lines or on the proximity to a 
certain galaxy type.  

\section*{Acknowledgments}
We gratefully acknowledge V. Mohan's help with the subtraction of QSO 
contribution from SDSS images and useful discussions with P. Serra.  
We thank the GMRT staff for their help during the observations.
GMRT is run by the National Centre
for Radio Astrophysics of the Tata Institute of Fundamental Research.
We acknowledge the use of SDSS spectra from the archive
(http://www.sdss.org/). RS and PPJ gratefully acknowledge support
from the Indo-French Centre for the Promotion of Advanced Research
(Project N.4304-2).

\def\aj{AJ}%
\def\actaa{Acta Astron.}%
\def\araa{ARA\&A}%
\def\apj{ApJ}%
\def\apjl{ApJ}%
\def\apjs{ApJS}%
\def\ao{Appl.~Opt.}%
\def\apss{Ap\&SS}%
\def\aap{A\&A}%
\def\aapr{A\&A~Rev.}%
\def\aaps{A\&AS}%
\def\azh{AZh}%
\def\baas{BAAS}%
\def\bac{Bull. astr. Inst. Czechosl.}%
\def\caa{Chinese Astron. Astrophys.}%
\def\cjaa{Chinese J. Astron. Astrophys.}%
\def\icarus{Icarus}%
\def\jcap{J. Cosmology Astropart. Phys.}%
\def\jrasc{JRASC}%
\def\mnras{MNRAS}%
\def\memras{MmRAS}%
\def\na{New A}%
\def\nar{New A Rev.}%
\def\pasa{PASA}%
\def\pra{Phys.~Rev.~A}%
\def\prb{Phys.~Rev.~B}%
\def\prc{Phys.~Rev.~C}%
\def\prd{Phys.~Rev.~D}%
\def\pre{Phys.~Rev.~E}%
\def\prl{Phys.~Rev.~Lett.}%
\def\pasp{PASP}%
\def\pasj{PASJ}%
\def\qjras{QJRAS}%
\def\rmxaa{Rev. Mexicana Astron. Astrofis.}%
\def\skytel{S\&T}%
\def\solphys{Sol.~Phys.}%
\def\sovast{Soviet~Ast.}%
\def\ssr{Space~Sci.~Rev.}%
\def\zap{ZAp}%
\def\nat{Nature}%
\def\iaucirc{IAU~Circ.}%
\def\aplett{Astrophys.~Lett.}%
\def\apspr{Astrophys.~Space~Phys.~Res.}%
\def\bain{Bull.~Astron.~Inst.~Netherlands}%
\def\fcp{Fund.~Cosmic~Phys.}%
\def\gca{Geochim.~Cosmochim.~Acta}%
\def\grl{Geophys.~Res.~Lett.}%
\def\jcp{J.~Chem.~Phys.}%
\def\jgr{J.~Geophys.~Res.}%
\def\jqsrt{J.~Quant.~Spec.~Radiat.~Transf.}%
\def\memsai{Mem.~Soc.~Astron.~Italiana}%
\def\nphysa{Nucl.~Phys.~A}%
\def\physrep{Phys.~Rep.}%
\def\physscr{Phys.~Scr}%
\def\planss{Planet.~Space~Sci.}%
\def\procspie{Proc.~SPIE}%
\let\astap=\aap
\let\apjlett=\apjl
\let\apjsupp=\apjs
\let\applopt=\ao
\bibliographystyle{aa}
\bibliography{/Users/neeraj/Desktop/mybib}

\end{document}